# Dual-Diffusion: Dual Conditional Denoising Diffusion Probabilistic Models for Blind Super-Resolution Reconstruction in RSIs

Mengze Xu, Jie Ma, Yuanyuan Zhu

*Abstract*—Previous super-resolution reconstruction (SR) works are always designed on the assumption that the degradation operation is fixed, such as bicubic downsampling. However, as for remote sensing images, some unexpected factors can cause the blurred visual performance, like weather factors, orbit altitude, *etc*. Blind SR methods are proposed to deal with various degradations. There are two main challenges of blind SR in RSIs: 1) the accurate estimation of degradation kernels; 2) the realistic image generation in the ill-posed problem. To rise to the challenge, we propose a novel blind SR framework based on dual conditional denoising diffusion probabilistic models (DDSR). In our work, we introduce conditional denoising diffusion probabilistic models (DDPM) from two aspects: kernel estimation progress and reconstruction progress, named as the dual-diffusion. As for kernel estimation progress, conditioned on low-resolution (LR) images, a new DDPM-based kernel predictor is constructed by studying the invertible mapping between the kernel distribution and the latent distribution. As for reconstruction progress, regarding the predicted degradation kernels and LR images as conditional information, we construct a DDPM-based reconstructor to learning the mapping from the LR images to HR images. Comprehensive experiments show the priority of our proposal compared with SOTA blind SR methods. Source Code is available at https://github.com/Lincoln20030413/DDSR

*Index Terms*—Blind super-resolution, diffusion model, remote sensing, deep learning

## I. INTRODUCTION

SUPER resolution (SR) aims to recover high-resolution (HR) images with more details from low-resolution (LR) images, which can restore many important details of remote sensing images (RSIs) degraded due to orbit altitude, weather factors and so on. SR is a highly ill-posed challenging task as different HR images can be mapped to the same LR image via different degradation models [1].

Recently, as the development of deep learning techniques, convolutional neural network (CNN)-based SR algorithms have become the mainstream while they have achieved satisfactory SR results [2]. However, most of the existing SR methods assume that LR images are generated from HR images by an ideal and fixed degradation model such as bicubic downsampling, which heavily limit the performance of those methods as the degradation in real scenes is mostly different from the ideal assumption. To depict the real scenes, an LR RSI $I_{LR}$ is degraded from an HR image $I_{HR}$ as [3]:

$$I_{LR} = (I_{HR} * \text{k}) \downarrow_s \qquad (1)$$

where $I_{HR}$ is the HR image, * denotes convolution operations of $I_{HR}$ with kinds of blur kernels k, $\downarrow_s$ denotes downsampling operations with a scale factor $s$. In this case, it is impossible to model every single degradation (e.g. bicubic downsampling) like previous methods because there exist huge kinds of degradation models for different LR images in real scenes.

To fill this gap, several studies have concentrated on using a single degradation-adaptive SR model without prior degradation information, known as blind SR, to cope with different degradations in real scenes. In blind SR tasks, the reconstruction process can be divided into two successive stages [6]: 1) degradation kernel estimation from the LR image and 2) kernel-based HR image reconstruction. Following the baseline, Luo *et al.* [6] proposed an end-to-end alternating single network (DAN) to estimate degradation kernel and reconstruct HR image iteratively. Luo *et al.* [4] proposed to reformulate the degradation model and apply the deep constrained least squares (DCLS) deconvolution module, achieving the state-of-the-art (SOTA) SR performance. However, both DAN and DCLS produce overly smoothing SR results.

To achieve realistic visual quality, other studies have introduced deep generative models including GANs and flow-based models in the blind SR problems [5][9], which have predicted sharper results with better perceptual quality. However, these models often suffer from the following limitations: The GAN-based models easily fall into mode collapse [8] and still neglect the ill-posed issue by only predicting one single SR result for one input. The flow-based models solve the ill-posed issue by adopting the invertible encoder to study the distribution instead of generating single result, but they endure architectural constraints and suboptimal sample quality [8].

To sum up, there are three main drawbacks in the existing blind SR models:

 1) **insufficient optimization object**: most of CNN-based blind SR algorithms are optimized by pixel-level losses [4][6], resulting in overly smoothing visual results which are not suitable for RSIs due to high-frequency information loss.

 2) **inaccurate kernel estimation:** it is difficult to perfectly estimate accurate degradation kernels directly from LR images due to the ambiguity generated by undersampling step [4].

This work was supported in part by the National Natural Science Foundation of China under Grant 62101052, *(Corresponding author: Jie Ma.)*
The authors are with the School of Information Science and Technology, Beijing Foreign Studies University, Beijing 100089, China (e-mail: majie_sist@bfsu.edu.cn).

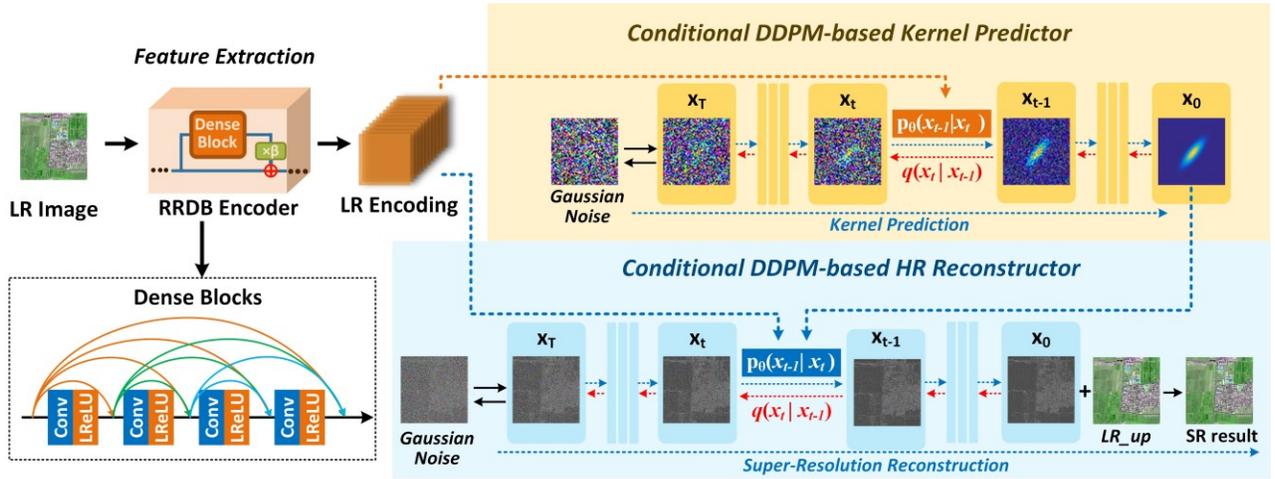

Fig. 1. Overall framework of the proposed method

**3) ill-posed problem:** on the ill-posed nature of SR problems, most of blind SR algorithms actually ignore it as they deterministically map from a LR image to a HR image.

Lately, Ho *et al.* [7] introduced denoising diffusion probabilistic models (DDPMs) into image synthesis. DDPMs are easy to define and efficient to train while they can also produce high-quality images [7]. More importantly, as the nature of generating data in complicated distribution, DDPMs can perfectly solve the ill-posed problem of SR.

Considering these, we propose a novel blind SR model (DDSR) based on dual denoising diffusion probabilistic model, which can successfully alleviate the aforementioned drawbacks. Firstly, we proposed a conditional DDPM-based kernel predictor to estimate the degradation kernel conditioned on the LR image. Then, a conditional DDPM-based HR reconstructor is built, where the predicted kernel and the LR image are regarded as the conditions to reconstruct the HR image. The main contributions of this letter are as follows:

1) We propose a novel DDPM-based blind SR framework for RSIs named DDSR, owning no architectural constraints like flow-based models and no mode collapse like GANs while producing more realistic SR results than the SOTA blind SR algorithms and DDPM-based non-blind SR method.
2) We propose a new DDPM-based kernel strategy to estimate accurate degradation kernel rather than directly estimate it from the LR images, which can be embedded in any other blind SR models.

## II. Proposed Method

This section first gives a brief introduction of DDPM, and then formally introduces our method which consists of two main components: a conditional DDPM-based kernel predictor and a conditional DDPM-based HR reconstructor. And more detailed information about the corresponding parameterization method, training phase and inference phase of our method follow after that.

### A. Denoising Diffusion Probabilistic Model

Diffusion model is a kind of generative model based on a parameterized Markov chain. It is trained using variational inference to gradually generate the sample $x_0 \sim p(x_0)$ in complicated distribution from a latent variable $x_T$ in simple distribution, which can be formulated as:

$$p_\theta(x_0) := \int p_\theta(x_{0:T}) dx_{1:T}, \quad (2)$$

where $p_\theta(x_{0:T})$ denotes the joint distribution of $x_0$ and latent variables $x_1, \cdots, x_T$, $\theta$ denotes the model parameters.

The diffusion model can be decomposed into two opposite processes: 1) the reverse process and 2) the diffusion process.

***Reverse process***: The computation progress of joint distribution $p_\theta(x_{0:T})$ under Markov chain, usually called the reverse process, converts the latent variable distribution $p_\theta(x_T)$ to the sample distribution $p_\theta(x_0)$:

$$p_\theta(x_{0:T-1} | x_T) := \prod_{t=1}^{T} p_\theta(x_{t-1} | x_t) \quad (3)$$

and the transition follows a Gaussian distribution, that is:

$$p_\theta(x_{t-1} | x_t) := N(x_{t-1}; \mu_\theta(x_t, t), \Sigma_\theta(x_t, t)), \quad (4)$$

where the diffusion timestep $t \in \{1, \cdots, T\}$ and the starting distribution $p(x_T) = N(x_T; 0, I)$

Following Ho *et al.* [7], the parameterization for the reverse process are as follows:

$$\mu_\theta(x_t, t) := \frac{1}{\sqrt{\alpha_t}} \left( x_t - \frac{\beta_t}{\sqrt{1-\bar{\alpha}_t}} \varepsilon_\theta(x_t, t) \right), \quad (5)$$

$$\Sigma_\theta(x_t, t) = \frac{1-\bar{\alpha}_{t-1}}{1-\bar{\alpha}_t} \beta_t I =: \sigma_t^2 I, \quad (6)$$

where $\varepsilon_\theta$ is a noise predictor, with the input of $x_t$ and $t$.

The optimization object can be formulated as:

$$\min_\theta L_s(\theta) = E_{x_0, \varepsilon, t} \left[ \left\| \varepsilon - \varepsilon_\theta \left( \sqrt{\bar{\alpha}_t} x_0 + \sqrt{1-\bar{\alpha}_t} \varepsilon, t \right) \right\| \right] \quad (7)$$

***Diffusion process:*** The approximation of posterior $q_\theta(x_{1:T} | x_0)$, which transforms the sample distribution $q_\theta(x_0)$ to the latent variable distribution $q_\theta(x_T)$, is the diffusion process. It is based on a Markov chain that gradually adds Gaussian noise to the sample according to a variance sequence $\beta_1, \cdots, \beta_T$, which can be formulated as:

$$q(x_{1:T} | x_0) := \prod_{t=1}^{T} q(x_t | x_{t-1}) \quad (8)$$

$$q(x_t | x_{t-1}) := N(x_t; \sqrt{1-\beta_t} x_{t-1}, \beta_t I), \quad (9)$$

where $\{\beta_t\}_{t=1}^T$ are constant hyperparameters.

Letting $\alpha_t = 1 - \beta_t$, $\alpha_t = 1 - \beta_t$, $\bar{\alpha}_t := \prod_{s=1}^t \alpha_s$, $x_t$ can be sampled at an arbitrary timestep $t$ in closed form:

$$q(x_t | x_0) = N(x_t; \sqrt{\bar{\alpha}_t} x_0, (1-\bar{\alpha}_t)I), \tag{10}$$

which can be reparameterized as:

$$x_t(x_0, \varepsilon) = \sqrt{\bar{\alpha}_t} x_0 + \sqrt{1-\bar{\alpha}_t} \varepsilon, \ \varepsilon \sim N(0, I) \tag{11}$$

*B. Conditional DDPM-based Kernel Predictor*

We construct a new conditional DDPM-based kernel predictor, which studies both the inherent nature of the kernel and information of LR images via the invertible mapping between the kernel distribution and latent variable distribution. Compared with the kernel estimation strategies in previous blind SR methods [4][5] which merely concentrate on LR images, our kernel predictor can predict more accurate kernel information. The key idea is to consider the LR images as the condition in reverse process to gradually predict the degradation kernel $x_0^{\text{ker}}$ from a latent variable $x_T^{\text{ker}}$ in definite distribution.

Letting $f_\theta$ denote the RDDB LR encoder and $u = f_\theta(\text{LR})$ denote LR encoding which contains content information of the LR image. The reverse process is defined as:

$$p_\theta^{\text{ker}}(x_0^{\text{ker}},...,x_{T-1}^{\text{ker}} | x_T^{\text{ker}}) := \prod_{t=1}^T p_\theta^{\text{ker}}(x_{t-1}^{\text{ker}} | x_t^{\text{ker}}, u) \tag{12}$$

$$p_\theta^{\text{ker}}(x_{t-1}^{\text{ker}} | x_t^{\text{ker}}) := N(x_{t-1}^{\text{ker}}; \mu_\theta(x_t^{\text{ker}}, t, u), \sigma_t^2 I) \tag{13}$$

where the initial distribution $p(x_T^{\text{ker}}) = N(0, I)$

According to Eq.(4)–(7), the reverse process is determined by the noise predictor $\varepsilon_\theta^{\text{ker}}$. Meanwhile, Eq.(11) shows that the **clean samples** will be accurately predicted if the noise predictor $\varepsilon_\theta^{\text{ker}}$ estimates noise precisely enough. The architecture of the noise predictor $\varepsilon_\theta^{\text{ker}}$ is the same as SRDiff [19].

*C. Conditional DDPM-based HR Reconstructor*

As for reconstruction part, we design a conditional DDPM-based framework to restore the difference $x^{img}$ between the real HR image and the upsampled LR image. Initially, the LR encoding $u$ and estimated kernel $v$ are separately fed as conditions in the transition of the reverse process to gradually predict the difference $x_0^{img}$ from a latent variable $x_T^{img}$ in definite distribution. Then, the HR image is reconstructed by adding the predicted difference $x_0^{img}$ to the upsampled LR image. Letting $g_\theta$ denote the kernel predictor that estimate degradation kernel of the LR image and thus the estimated kernel $v = g_\theta(x_{LR})$, the reverse process is defined as:

$$p_\theta^{img}(x_0^{img},...,x_{T-1}^{img} | x_T^{img}) := \prod_{t=1}^T p_\theta^{img}(x_{t-1}^{img} | x_t^{img}, u, v) \tag{14}$$

$$p_\theta(x_{t-1}^{img} | x_t^{img}) := N(x_{t-1}^{img}; \mu_\theta(x_t^{img}, t, u, v), \sigma_t^2 I) \tag{15}$$

where the initial distribution $p(x_T^{img}) = N(0, I)$

Here, we introduce attention over convolution operation and construct a new dynamic convolutional noise predictor $\varepsilon_\theta^{img}$ to increase the representation capability. As shown in Fig. 2, we use U-Net [10] as the backbone for the noise predictor. For the main body, we first take a convolution block which consists of a convolutional layer and Mish activation [11] to extract shallow features from $x_t^{img}$. Then the extracted features are fused with the LR image encoding $u$ and fed into the following network.

In the U-Net, the compression path and the expansion path both consist of four feature extraction groups, each of which has two dynamic residual blocks and a downsampling/upsampling layer. The downsampling layer in the compression path is two-stride convolution and the upsampling layer in the expansion path is transposed convolution. Inserted between the two paths, the middle step is composed of two residual blocks. Each of the dynamic residual blocks consists of two dynamic convolutional layers [12].

To make good use of the information of timestep and the predicted kernel, each dynamic residual block accepts a timestep encoding and an estimated kernel $v$ as conditions. Following Ho *et al.* [7], the timestep encoding is the result of the diffusion timestep $t$ transformed by the transformer sinusoidal positional encoding [13]. Eventually, another convolution block is adopted to produce noise $\varepsilon_{t-1}^{img}$, which is then used to restore $x_{t-1}^{img}$ according to Eq.(4), (5), (6).

*D. Training and Inference*

---

**Algorithm 1** Training

**Input:** Total diffusion timesteps $T$, paired LR-HR image patches $\{(x_{LR}, x_{HR})\}_{k=1}^K$, hyperparameters $\bar{\alpha}_t$, the corresponding degradation kernels $\{x^{\text{ker}}\}_{k=1}^K$, pretrained LR encoder $f_\theta$

1: **Initialize:** initialized noise predictor $\varepsilon_\theta^{\text{ker}}$ of the kernel predictor
2: **repeat**
3:   Sample $x^{\text{ker}}$
5:   Compute the LR encoding $u = f_\theta(x_{LR})$
6:   Randomly sample $\varepsilon \sim N(0, I)$, $t \sim U(\{1,...,T\})$
7:   According to Eq. (7), (11), (13), Take a gradient step on
     $\nabla_\theta \left\| \varepsilon - \varepsilon_\theta^{\text{ker}}\left(x_t^{\text{ker}}, u, t\right) \right\|$, $x_t^{\text{ker}} = \sqrt{\bar{\alpha}_t} x^{\text{ker}} + \sqrt{1-\bar{\alpha}_t} \varepsilon$
8: **until** converged
9: **Initialize:** initialized noise predictor $\varepsilon_\theta^{img}$ of the reconstructor, the trained kernel predictor $g_\theta$
10: **repeat**
11:   Sample $(x_{LR}, x_{HR})$
12:   Upsample $x_{LR}$ as $up(x_{LR})$, compute $x^{img} = x_{HR} - up(x_{LR})$
13:   Compute the LR encoding $u = f_\theta(x_{LR})$
14:   Predict the kernel $v = g_\theta(x_{LR})$
15:   Randomly sample $\varepsilon \sim N(0, I)$, $t \sim U(\{1,...,T\})$
16:   According to Eq. (7), (11), (15), Take a gradient step on
     $\nabla_\theta \left\| \varepsilon - \varepsilon_\theta^{img}\left(x_t^{img}, u, v, t\right) \right\|$, $x_t^{img} = \sqrt{\bar{\alpha}_t} x^{img} + \sqrt{1-\bar{\alpha}_t} \varepsilon$
17: **until** converged

---

In the training phase, illustrated in Algorithm 1, we successively optimize the noise predictors of kernel predictor and the reconstructor. The former is done conditioned on the LR images and corresponding degradation kernels. Then, the latter is done conditioned on the LR images and the kernels predicted by the former trained kernel predictor.

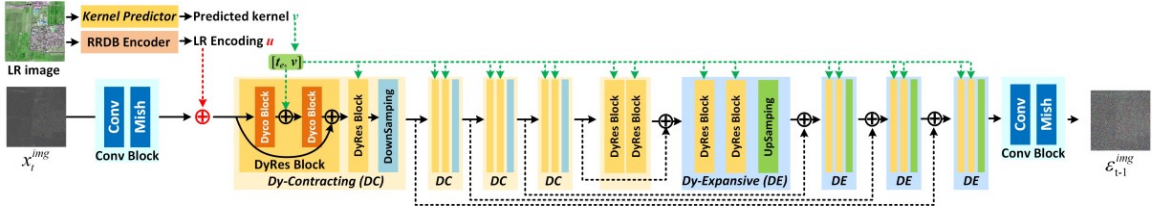

Fig. 2. The architecture of the noise predictor for the HR reconstructor

The inference phase is illustrated in Algorithm 2. Firstly, with LR encoding $u$ as the condition, we start from a latent variable $x_T^{ker}$ in the standard Gaussian distribution, gradually predict and remove the noise, and finally gain the predicted kernel $x_0^{ker}$. After that, with $u$ and $x_0^{ker}$ as conditions, we start from a latent variable $x_T^{img}$ in the standard Gaussian distribution, iteratively predict and remove the noise, and finally get the predicted difference. The final SR result is gained by adding the difference on the upsampled LR image.

---

**Algorithm 2** Inference

**Input:** Total diffusion timesteps $T$, LR image $x_{LR}$
**Load:** pretrained $f_\theta$, trained $\varepsilon_\theta^{ker}$ and $\varepsilon_\theta^{img}$
1: Sample $x_T^{ker}, x_T^{img} \sim N(0,1)$
2: Compute the LR encoding $u = f_\theta(x_{LR})$
3: Upsample $x_{LR}$ as $up(x_{LR})$
4: **for** $t = T, T-1, ..., 1$ **do**
5:     Sample $z \sim N(0,I)$ if $t > 1$, else $z = 0$
6:     Compute $x_{t-1}^{ker}$ by Eq. (4), (5), (6), (10):
$$x_{t-1}^{ker} = \frac{1}{\sqrt{\alpha_t}}\left(x_t^{ker} - \frac{1-\alpha_t}{\sqrt{1-\bar{\alpha}_t}}\varepsilon_\theta^{ker}(x_t^{ker},u,t)\right) + \sigma_t^2 Iz$$
7: **end for**
8: Take $x_0^{ker}$ as the predicted kernel $v$
9: **for** $t = T, T-1, ..., 1$ **do**
10:     Sample $z \sim N(0,I)$ if $t > 1$, else $z = 0$
11:     Compute $x_{t-1}^{img}$ by Eq. (4), (5), (6), (10):
$$x_{t-1}^{img} = \frac{1}{\sqrt{\alpha_t}}\left(x_t^{img} - \frac{1-\alpha_t}{\sqrt{1-\bar{\alpha}_t}}\varepsilon_\theta^{img}(x_t^{img},u,v,t)\right) + \sigma_t^2 Iz$$
12: **end for**
13: Take $x_0^{img}$ as the predicted difference $x_0^{img}$
14: **return** $x_0^{img} + up(x_{LR})$ as SR result $x_{SR}$

---

## III. EXPERIMENTS

In this section, we first introduce the experimental settings including datasets, model configurations and details in training and evaluation. Then, we demonstrate the experimental results.

### A. Datasets

To evaluate the effectiveness of our method, we employ RSIs offered by GeoEye-1 satellite and GoogleEarth. Here, 120 of the GeoEye-1 dataset containing 130 multispectral RSIs with 0.41m-resolution and 229 of the GoogleEarth dataset containing 239 optical RSIs with 1m-resolution are used for training while the remaining 20 images are used for testing. Then, according to Eq.(1), we synthesize corresponding LR images with specific degradation settings (bicubic downsampling and kinds of anisotropic Gaussian kernels). Here, following [9], we choose more general degradation kernels: the anisotropic Gaussian kernels. The size of the kernels is fixed at $24 \times 24$ while the kernels are determined by the covariance matrix. The covariance matrix is determined by two random eigenvalues $\lambda_1, \lambda_2 \sim U(0.2, 4)$ and a random rotation angle $\theta \sim U(0, \pi)$.

### B. Implementation Details

In the diffusion models, the diffusion steps $T$ is set to 100 and the noise variance $\beta_1, \cdots, \beta_T$ is defined by the setting in [19]. The channel numbers of the convolutional layers in the noise predictors are all set to 64. For the LR encoder, the numbers of RRDBs are set to 8 and the channel numbers are set to 32.

The proposed DDSR is trained on the PC equipped with an Intel Core i9-10940X and a GPU NVIDIA Geforce RTX 3090. Besides the commonly used evaluation metric PSNR, we also evaluate our method on FID [14] and LPIPS [15], which can better measure the reality and visual quality of SR results. The lower FID, the better the model generation results on image diversity and reality. The lower LPIPS, the better the model generation results on human visual perception.

### C. Experimental Results

We first compare our method with recent SOTA blind SR methods, DANv1 [6], DASR [16], DCLS [4], which are optimized by pixel-level losses. Also, we choose two pioneer kernel estimation methods, which are GAN-based KernelGAN [5] and flow-based DIP-FKP [9]. Following their proposers, we combine them with SOTA non-blind SR methods including ZSSR [17] and USRNet [18] respectively, as blind SR algorithms. What's more, the pioneer DDPM-based non-blind SR method SRDiff [19] and bicubic upsampling are taken for comparison.

Table I shows the comparison of objective metrics with different blur kernel parameters including two random eigenvalues $\lambda_1$, $\lambda_2$ and a random rotation angle $\theta$. The DANv1, DASR, DCLS optimized by pixel-level loss all achieve high PSNR, but perform poorly on visual metrics LPIPS and FID. The DDPM-based SRDiff performs better than the aforementioned algorithms on LPIPS and FID. However, the KernelGAN and DIP-FKP both perform badly on three metrics. The proposed DDSR outperforms all competitive algorithms in terms of FID and LPIPS while also outperforms the KernelGAN, DIP-FKP and SRDiff on PSNR. Fig. 3 shows the visual comparison, confirming the above objective indications. Our proposed DDSR generate SR results which are more realistic and own more texture details.

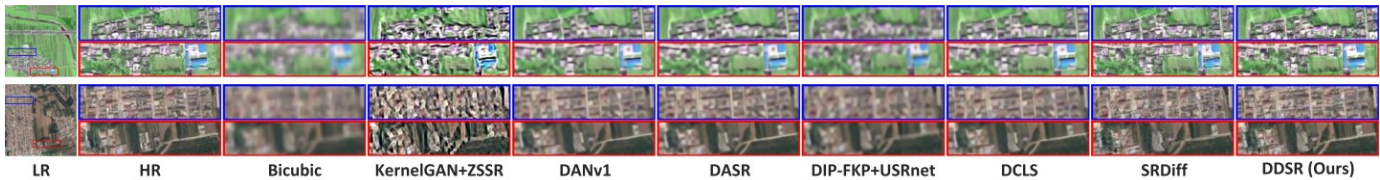

Fig. 3. The top and the bottom rows are ×4 SR results from GeoEye-1 and GoogleEarth dataset with a blur kernel of $\lambda_1=1.2$, $\lambda_2=2.4$, $\theta=0$

TABLE I
QUANTITATIVE COMPARISON OF ×4 SR RESULTS ON THE GEOEYE-1 DATASET AND THE GOOGLEEARTH. THE BEST RESULTS ARE IN **BOLD**

| | | $[\lambda_1, \lambda_2, \theta] = [1.2, 2.4, 0]$ | | | $[\lambda_1, \lambda_2, \theta] = [1.2, 2.4, \pi/4]$ | | | $[\lambda_1, \lambda_2, \theta] = [3.6, 2.4, 0]$ | | |
|---|---|---|---|---|---|---|---|---|---|---|
| | | PSNR | LPIPS | FID | PSNR | LPIPS | FID | PSNR | LPIPS | FID |
| Bicubic | GeoEye | 21.73 | 0.687 | 280.73 | 21.82 | 0.701 | 273.53 | 21.11 | 0.816 | 329.40 |
| KernelGAN[5]+ZSSR[17] | | 18.70 | 0.486 | 270.87 | 18.16 | 0.514 | 236.10 | 21.75 | 0.464 | 243.46 |
| DANv1[6] | | 23.65 | 0.439 | 178.82 | 23.68 | 0.431 | 176.66 | 23.36 | 0.471 | 213.34 |
| DASR[16] | | 23.46 | 0.477 | 197.53 | 23.49 | 0.468 | 200.45 | 23.18 | 0.504 | 235.75 |
| DIP-FKP[9]+USRnet[18] | | 18.63 | 0.617 | 237.53 | 18.41 | 0.610 | 261.32 | 18.85 | 0.717 | 301.31 |
| DCLS[4] | | **23.71** | 0.430 | 178.55 | **23.76** | 0.421 | 187.28 | **23.43** | 0.463 | 207.83 |
| SRDiff[19] | | 21.57 | 0.182 | 90.89 | 21.55 | 0.180 | 102.15 | 21.20 | 0.205 | 118.95 |
| DDSR (Ours) | | 21.74 | **0.169** | **81.46** | 21.74 | **0.172** | **86.27** | 21.32 | **0.194** | **106.36** |
| Bicubic | Google | 23.82 | 0.604 | 192.39 | 23.79 | 0.626 | 194.54 | 22.73 | 0.718 | 229.97 |
| KernelGAN[5]+ZSSR[17] | | 20.00 | 0.472 | 227.32 | 20.07 | 0.366 | 237.14 | 23.07 | 0.453 | 209.35 |
| DANv1[6] | | 26.38 | 0.368 | 178.46 | 26.33 | 0.365 | 189.18 | 25.95 | 0.393 | 184.38 |
| DASR[16] | | 26.16 | 0.373 | 165.32 | 26.13 | 0.375 | 171.75 | 25.68 | 0.399 | 184.94 |
| DIP-FKP[9]+USRnet[18] | | 19.72 | 0.511 | 200.32 | 19.53 | 0.545 | 193.46 | 19.81 | 0.510 | 259.72 |
| DCLS[4] | | **26.42** | 0.355 | 171.70 | **26.40** | 0.356 | 175.80 | **25.98** | 0.382 | 177.87 |
| SRDiff[19] | | 23.81 | 0.185 | 86.10 | 23.81 | 0.196 | 83.01 | 23.35 | 0.207 | 85.53 |
| DDSR (Ours) | | 24.26 | **0.176** | **80.99** | 24.22 | **0.183** | **72.29** | 23.75 | **0.197** | **79.76** |

## IV. CONCLUSION

In this letter, we have proposed a novel blind SR method based on dual conditional denoising diffusion probabilistic models. Firstly, we propose a conditional DDPM-based kernel predictor to estimate the degradation kernel conditioned on LR images. Then, a conditional DDPM-based HR reconstructor is constructed to recover the HR images conditioned on the estimated kernels and LR images. Experimental results shows that the proposed algorithm outperforms the SOTA algorithms in terms of the reality and visual quality of SR results. In the future, we will explore our model's adaptation to more diverse degradation scenes.